\documentclass[12pt]{iopart}
\usepackage{longtable}
\usepackage{xcolor}
\usepackage{graphicx}
\pdfoutput=1
\begin{document}
\title[Beyond-DFT Studies of Thiophene-Furan oligomers]{Thiophene-Furan oligomers: Beyond-DFT Study of Electronic and Optical Properties}
\author{V A Bastos, T J da Silva and M J Caldas}
\address{Institute of Physics, University of São Paulo, São Paulo, Brazil}
\eads{\mailto{vinicius.alves.bastos@usp.br},\mailto{talesjds@usp.br},\mailto{mjcaldas@usp.br}}

\begin{abstract}

Thiophene oligomers are an important class of organic materials for photovoltaic applications, owing to their unique optoelectronic properties.  Recently it was suggested that incorporation of furan units to the thiophene chains, maintaining the chain structure, namely thienylfuran linear oligomers, can bring improvements to the final material. In this work, we present a theoretical study of thiophene, furan and thienylfuran short chains, up to 4 units. Structural and electronic properties were obtained using Hartree-Fock (HF) and Density Functional Theory (DFT) calculations plus beyond mean-field methodologies, specifically Second-order M{\o}ller-Plesset perturbation theory on HF (HF-MP2) and many-body perturbation theory by the G0W0 approximation on DFT (G0W0@DFT). The optical properties were calculated on top of G0W0@DFT data using the Bethe-Salpeter Equation. We investigate properties from the monomers \textit{T} and \textit{F} to tetramers with different sequencing of units \textit{TT}, \textit{TF} or \textit{FF}, always bonded through the usual carbon atoms. As well known for the uniform oligothiophene chains, also here for any \textit{TT} sequencing we find a torsion angle of $150^{\circ}$, while all other sequencing of units result planar, which can be relevant for film producing. Also we find that the first optical transitions for the oligomers reaches a promising threshold of $\sim 3$ eV at the tetramer length, combined with ionization potentials around $\sim 7$ eV, which confirms the relevance of these organic compounds for photovoltaic applications.

\end{abstract}

\noindent{\it Photovoltaics, organic molecules, GW, Bethe-Salpeter \/}

\maketitle

\section{Introduction}

Thiophene (\textit{T}) polymers or oligomers are a common choice for the active layer of hybrid organic-inorganic solar cells, given the special qualities regarding optical absorption energies, and convenient alignment of ionization potentials with band gaps of consolidated semiconducting oxides \cite{kim+03nl,blum+08apl,sant+14jpcc,bell+94jpc,filh+07ptrs,holl+14pccp}. Oligothiophenes films can be built from clean oligomers \cite{bhag+21es}, or most often from alkyl-functionalized units which, depending on the specific preparation method, can either bring disorder \cite{ramo-cald18arxiv} or be used to neatly arrange \cite{sney+21sa} the film. In any case, the final result is very stable.

Concerning furan (\textit{F}) oligomers, highly soluble in common organic solvents and that can be processed without alkylation, regarding optoelectronic properties are comparable to oligothiophenes \cite{gand11gc,gidr+10jacs,gidr+11cc,gidr-bend14ac}, however for photovoltaic devices there may be issues related to instability in the presence of light and oxygen\cite{cao-rupa17cej}.

Recently, a new class of polymers from thiophene-furan combinations has emerged, addressing some of the issues discussed above concerning preparation, conductivity and light interaction,\cite{gidr+13cc,xion+16am,yuny+16acsml,stee+19jpcc,yao+20mcf,yao+21CJPS} which point to a favourable behavior compared to the original thiophene or furan oligomers. With the motivation of studying these combinations, we present here a theoretical study of structural, electronic and optical properties of finite isolated thiophene-furan oligomers of up to 4 units, in all cases using ab-initio methods starting from Density Functional Theory for structural properties, and going to Many Body Perturbation Theory and the Bethe-Salpeter formalism for electronic and optical properties. We find that indeed thienylfuran oligomers promise to combine the best properties of the original moieties. 

\section{Methodology}

We here briefly describe the methodology, in the following sections we show and discuss our results for the structural properties of different \textit{T}-\textit{F} combinations, move to energy alignment and  optical properties.

All  initial structural results were obtained using the NWChem code \cite{vali+10cpc} with atom-centered basis sets.  In order to obtain the bithiophene \textit{2T}, bifuran \textit{2F} and thienylfuran \textit{TF} conformations and torsion potentials, the geometries were optimized by adopting as a first step Density Functional Theory (DFT) \cite{hohe+64pr,kohn+65pr} at the PBE0/cc-pVDZ \cite{perd+96prl,perd+96jcp,adam-baro99jcp,dunn89jcp} level. All optimization processes took into account constraints by keeping fixed point group symmetries, where applicable, and fixed \textit{X}-\textit{C}-\textit{C}-\textit{X} dihedral angles, \textit{X} being the heteroatoms \textit{S} or \textit{O}. On top of the optimized geometries, with the same basis set, we performed non-relax M{\o}ller-Plesset perturbation theory on Hartree-Fock (HF-MP2) calculations to extract the total energies (as pointed in previous works \cite{raos+03cpl,bloo+14jctc}, HF-MP2 methods are more reliable than DFT in describing the energy profile of the bithiophene conformation). In an effort to confirm our results for the torsion potentials, for the special angles $0^{\circ}$, $40^{\circ}$, $90^{\circ}$, $150^{\circ}$ and $180^{\circ}$ we performed complete HF-MP2 relaxations, for all dimers, keeping fixed the angles and the relevant symmetry constraints.

The dimers mimimum-energy geometries serve as the starting point for building the related trimers and tetramers. The geometries in this case were optimized keeping the relevant symmetry constraints through PBE0, 6-31g(d,p) basis set; in the special case of \textit{3T} we kept fixed the dihedral angles at $150^{\circ}$. 

The electronic and optical properties were calculated for the optimized geometries using the Fritz Haber Institute ab initio molecular simulations package (FHI-aims) \cite{blum+09cpc,liu+20jcp}. We obtained the ground-state properties with the PBE0 functional, NAO-VCC-3Z \cite{zhan+13njp} basis set, and we used the electronic charge density to calculate the atomic Hirshfeld charges \cite{hirs77tca}; further, we used the dimers atomic charges to evaluate their Molecular Electrostatic Potential \cite{jane-pete11cms} (MEP) over the Van der Waals surface \cite{bond64jpc} using the Jmol code \cite{jmol-jmol13}. The optical properties were calculated through the Bethe-Salpeter Equation (BSE) \cite{hedi65pr,hybe-loui85prl,godb+86prl,rohl-loui00prb} with the Tamm-Dancoff approximation, using the G0W0@PBE0 results as the starting point to obtain the singlet optical transitions. 

For the ionization potential (IP), for each system we tuned the percentage of exact-exchange ($\alpha$) of the PBEh (PBE$\alpha$) functional following the internally consistent procedure described in the work by Pinheiro \textit{et al.} \cite{pinh+15prb}, here also using the NAO-VCC-3Z basis set at the FHI-aims code. Basically, we look for minimizing the G0W0 quasi-particle correction for the Highest Occupied Molecular Orbital (HOMO) in order to get the IP.

\section{Results and Discussion}

\subsection{Structural Properties}

\begin{figure}
  \includegraphics[width=1\linewidth]{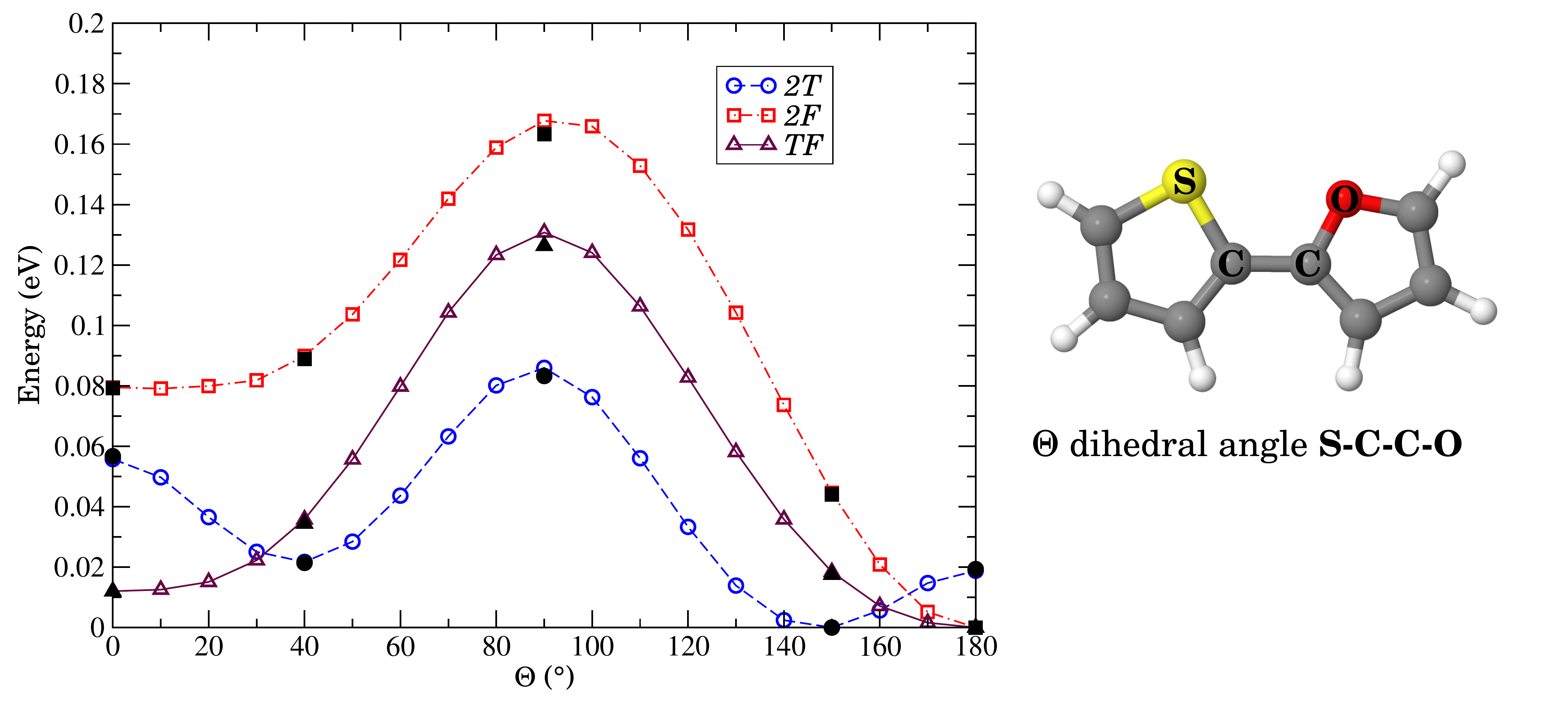}
  \caption{\label{torsion_potential}Torsion Potential for the \textit{2T}, \textit{TF} and \textit{2F} dimers. Open symbols: geometries optimized with PBE0 and energies evaluated using HF-MP2. Solid symbols: geometries and energies evaluated using HF-MP2. Symmetry constraints adopted where applicable. The angle is exemplified for the \textit{TF} dimer at $\Theta=0$.}
\end{figure}

For any sort of oligomer, and in particular for linear chains, the inter-ring torsion angles are key parameters, certainly relevant for the structural properties of related films. In Figure \ref{torsion_potential} we show the torsion potentials for the \textit{2T}, \textit{2F} and \textit{TF} dimers. It is worth to emphasize that the geometries obtained with PBE0 (open symbols) and HF-MP2 (solid symbols) are very similar: in terms of bond length, the largest differences are around hundredths of angstroms, and for internal angles at most tenths of degrees.

The \textit{2T} global minimum shows an \textit{antiparallel} torsion angle conformation at $150^{\circ}$, and a second minimum is obtained in a \textit{parallel} conformation at a torsion angle of $40^{\circ}$ (as seen previously \cite{raos+03cpl,bloo+14jctc} these results strongly depend on going to MP2 since the corresponding bare PBE0 angles are $160^{\circ}$ and $30^{\circ}$). Interestingly, the \textit{2F} and \textit{TF} global minima are found in antiparallel planar conformation, and the second minima are in parallel planar conformation. For the \textit{2F} dimer there is an extremely small difference of $\sim 10^{-4}${eV} indicating a minimum at $\sim 10^{\circ}$ for the parallel conformation, in close agreement with theoretical studies adopting full MP2 relaxation from Perkins et al\cite{perk+21JPCA}, where it is also reported the planarity of TF using a large variety of DFT and wavefunction approaches. Our results are thus in good accord with previous works \cite{raos+03cpl,bloo+14jctc,garc-karp09cpl,perk+21JPCA}. In particular for \textit{2T}, Samdal \textit{et al.} \cite{samd+93sm} found from gas phase experiments the existence of the two conformations, parallel at $36^{\circ}$, and antiparallel at $148^{\circ}$.

Other important points are the energetic barrier heights at $90^{\circ}$. The \textit{2F} potential shows a torsion barrier from antiparallel to parallel (ap-p) conformation of $6.6$ $k_BT$ (assuming $T=298$K), and a reverse barrier of $3.5$ $k_BT$. Consequently, at ambient conditions, the transition from an antiparallel to parallel configuration is very unfavourable, and we expect that the antiparallel-planar conformation will be the most probable. Turning now to the \textit{TF} dimer, it displays an ap-p barrier of $5.09$ $k_BT$ and the reverse very close in energy, $4.62$ $k_BT$. In this case, both barrier heights are quite high, then transitions from both configurations will be very unfavourable. We thus decided specifically for this intriguing case to investigate both the parallel-planar (\textit{TF$_p$}) and antiparallel-planar (\textit{TF$_{ap}$}) conformers. On the other hand, the \textit{2T} dimer has an ap-p  barrier height of $3.34$ $k_BT$ and the reverse in a very similar -but smaller- energy  of $2.56$ $k_BT$; both barriers are not so high, and transitions in both directions are allowed.  As expected from these results, at ambient conditions \textit{2T} is more commonly found in the non-planar antiparallel conformation, with an abundance of $70\%$ detected in NMR measurements \cite{vera+74jacs}.

\begin{figure}
  \includegraphics[width=1\linewidth]{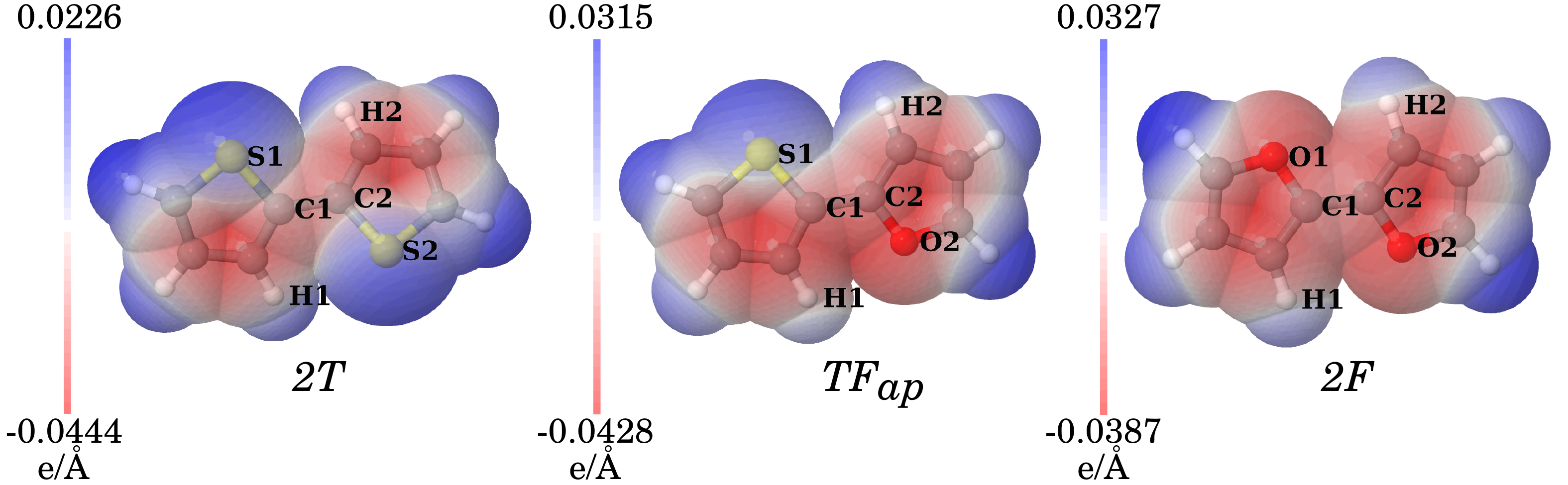}
  \caption{\label{meps}Molecular Electrostatic Potential (MEP) map for the \textit{2T}, \textit{TF$_{ap}$} (antiparallel conformation) and \textit{2F} optimal geometries.}
\end{figure}

The torsion potential curves give a clear indication of the preferable dimer conformations, and as an illustrative view we show in Figure \ref{meps} a purely electrostatic interpretation of these results, starting from fractional atomic charges obtained through DFT,  that is, the MEP graphs for the dimers. As a first comment, we see that as expected the \textit{S}-atoms are ``positively charged'' while the \textit{O}-atoms carry a ``negative charge''. We can note that the interaction between positively-charged sulfur and hydrogen atoms in the neighbor rings is enough to explain the \textit{2T} non-planarity. In the same direction, the attractive interaction between the opposite charges of oxygen and hydrogen atoms in the  \textit{2F} and \textit{TF} dimers will lead them to planarity -- however, the \textit{TF} dimer in any stable conformation always experience a repulsive interaction between atomic positive charges (\textit{S}-\textit{H} or \textit{H}-\textit{H})  in opposite rings. This explains why, as seen above, there is no significant difference in energy between the parallel-planar and the antiparallel-planar conformations. This result is very significant as we will comment below, and indeed even for more complex molecules near-planarity of \textit{TF} segments have been found in a recent theoretical study\cite{pach+20jmcc}. 

For the longer oligomers, the search for optimal geometries followed the same path, starting with torsion angles from the results for dimers at the DFT level, and proceeding to full-optimization at the HF-MP2 level at special configurations. As seen in Table \ref{torsion-angles} \textit{the results confirm the planarity of all structures in which the ordering does not include thiophene near-neighbours,  e.g. planarity of  \textit{TFT}, \textit{3F}, \textit{TFFT} and non-planarity of \textit{3T} and \textit{FTTF}}. This characteristic will be very important in the formation of films, since these planar systems could be packed more neatly, possibly even providing better $\pi$ stacking than thiophene oligomers.

\begin{table}[h!]
\caption{\label{torsion-angles}Dihedral angles of thiophene and furan oligomers and co-oligomers. Angles obtained from symmetry-constrained (except for \textit{TF} and \textit{TFTF}) optimizations through NWChem using PBE0 (values in parentheses HF-MP2).}
\begin{indented}
\item[]\begin{tabular}{@{}llll}
\br
 System & \textit{T}-\textit{T} & \textit{F}-\textit{F} &  \textit{T}-\textit{F} \\
\mr
 \textit{2T} & $157^{\circ}$ ($147^{\circ}$) & - & - \\
 \textit{2F} & - & $180^{\circ}$ ($180^{\circ}$) & - \\
 \textit{TF} & - & - & $180^{\circ}$ ($180^{\circ}$) \\
 \textit{3T} & $159^{\circ}$ ($150^{\circ}$) & - & - \\
 \textit{3F} & - & $180^{\circ}$ & - \\
 \textit{TFT} & - & - & $180^{\circ}$ \\
 \textit{FTF} & - & - & $180^{\circ}$  \\
 \textit{TFTF} & - & - & $180^{\circ}$ \\
 \textit{TFFT} & - & $180^{\circ}$ & $180^{\circ}$ \\
 \textit{FTTF} & $164^{\circ}$ ($154^{\circ}$) & - & $179^{\circ}$ ($177^{\circ}$)\\
\br
\end{tabular}
\end{indented}
\end{table}

\subsection{Optical Properties and Ionization Potential}

\begin{figure}
  \includegraphics[width=0.95\linewidth]{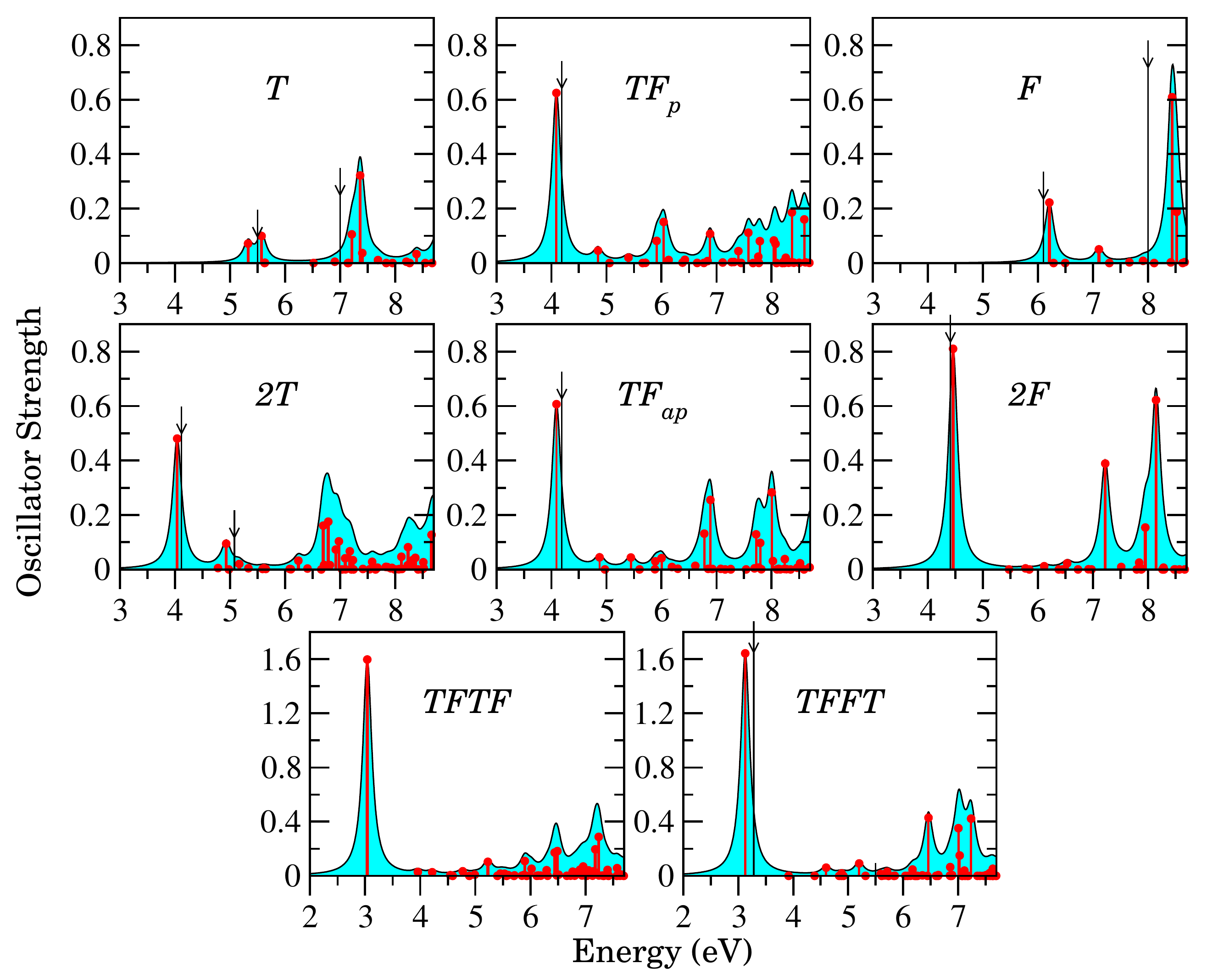}
  \caption{\label{absorption_spectra}Absorption spectra of thiophene \textit{T}  and furan \textit{F} oligomers obtained from BSE calculations. Red dots correspond to the excitation energies, and the red lines their respective oscillator strength. Blue solid curve is our estimate for the absorption spectra using Lorentzian functions with a scale parameter of 0.1 eV. Arrows indicate the experimental peak positions included in Table \ref{optical_transitions}.}
\end{figure}

We now present our results for the absorption properties of a subset including monomers, dimers, trimers and tetramers, in this case as indicated in Section 2 adopting the PBE0 functional, that properly describes the screening for the neutral molecules. We show in Figure \ref{absorption_spectra} the absorption spectra obtained for monomers and dimers, where we also indicate (black arrows) the experimental results for the more pronounced absorption peaks in this range, included for reference in Table \ref{optical_transitions}. We can see that our calculations show good agreement with experimental data \cite{bell+94jpc,holl+14pccp,palm+95cp}, not only for the excitation energies but also qualitatively for relative peak intensities.  Concerning theoretical studies, for the single thiophene molecule our results are in good accord with the BSE@GW perturbative and self-consistent approaches presented by Hung \textit{et al.} \cite{hung+16prb}. Furthermore, our results follow the same trends of beyond-Hartree-Fock methods in terms of the optical transition ordering and oscillator strengths discussed by Prlj \textit{et al.} \cite{prlj+15jpcl}. The main character of the lowest energy absorption peak is given by the HOMO$\rightarrow$LUMO transition for all systems, and for thiophene there is also a contribution of the HOMO-1$\rightarrow$LUMO at the lowest energy value in the absorption range. The single furan molecule is included in a test set of molecules (Thiel benchmark set) for theoretical results\cite{schr+08jcp}, and also in this case our result deviates just $0.3\%$ from the best estimate value.

The relevant transitions are of $\pi-\pi$* type, and as expected we see a redshift in the first peak when moving from monomers to dimers. An interesting behavior concerning the amplitude of the optical transitions, seen clearly in Fig.\ref{absorption_spectra}, is that while for the monomers the oscillator strengths increase strongly from the lowest transition ($\sim$5 eV) to higher energies, for the dimers we see the most intense absorption at the first transition at $\sim$4 eV.   

\begin{table}
\caption{\label{optical_transitions}BSE@G0W0 singlet optical excitations for thiophene and furan oligomers and co-oligomers studied here. All transitions are of $\pi \rightarrow \pi*$ character.}
\begin{indented}
\item[]\begin{tabular}{@{}llll}
\br
System & Energy (eV) & Osc. Strength (a.u.) & Exp. (eV)\\
\mr
\textit{T} & 5.33 & 0.071 & \\
 & & & 5.50 \cite{holl+14pccp} \\
 & 5.57 & 0.099 & \\
\mr
\textit{F} & 6.21 & 0.222 & 6.10 \cite{palm+95cp}\\
\mr
\textit{2T} & 4.04 & 0.481 & 4.09,4.12,4.29 \cite{gomb+04jpsj,diaz+81jecie,bell+94jpc} \\
\mr
\textit{TF$_{ap}$} & 4.09 & 0.607 & 4.19 \cite{wynb+71joc} \\
\mr
 \textit{TF$_p$} & 4.09 & 0.625 & 4.19 \cite{wynb+71joc} \\
\mr
\textit{2F} & 4.46 & 0.810 & 4.41 \cite{grig+66jcsc} \\
\mr
\textit{3T} & 3.41 & 0.989 & 3.50 \cite{gomb+04jpsj} \\
\mr
\textit{TFT} & 3.41 & 1.062 & - \\
\mr
\textit{FTF} & 3.44 & 1.130 & - \\
\mr
\textit{3F} & 3.70 & 1.198 & - \\
\mr
\textit{TFTF} & 3.04 & 1.597 & - \\
\mr
\textit{TFFT} & 3.13 & 1.643 & 3.28 \cite{gidr+13cc} \\
\br
\end{tabular}
\end{indented}
\end{table}

Focusing on \textit{TF}, it is important to emphasize the very promising optical properties we see in the absorption spectra at both conformations. The first peak is close to that of \textit{2T}, and additionally, we also find absorption in the range around 5 and 6 eV; these absorption peaks are not present in the \textit{2F} spectrum and are seen for the \textit{2T} that is, in \textit{TF} qualities from the monomers are combined. 

For longer oligomers, we show in Fig. \ref{absorption_spectra} the absorption spectra for \textit{TFTF} and  \textit{TFFT}, both planar; we see again higher intensity for the first transition, and the important property is the strong redshift in the first peak from dimers to tetramers, that brings now the onset of absorption to the limit UV-V between ultra-violet and visible light. 

For the estimation of the IP, we used a many-body corrected methodology \cite{pinh+15prb} specifically directed to obtain the removal-energy of the HOMO-orbital that obeys Koopmans' theorem, namely we adjust the percentage of exact-exchange ($\alpha$) present in hybrid PBE$\alpha$ in order to guarantee that the value obtained is not strongly changed when we apply the G0W0 corrections. We searched for the optimal values $\alpha_{ic}$ within a wide range of $\alpha$ values.

\begin{table}
\caption{\label{thio-furan_IP-alphaic}Ionization potential (IP) for thiophene and furan oligomers and co-oligomers studied here calculated with the indicated percentage of exact exchange $\alpha_{ic}$. Direct PBE$\alpha_{ic}$ value from the highest occupied orbital HOMO level (Koopmans), and from the subsequent G0W0 correction. Experimental data from the references indicated.}
\begin{indented}
\item[]\begin{tabular}{@{}lllll}
\br
 System & $\alpha_{ic}$ & Koopmans & G0W0 & Exp.\\
 &  & (eV) & (eV) & (eV)\\
\mr
\textit{T} & $76\%$ &  9.01 & 9.01 & 8.85 \cite{klas+82jcs}\\
\mr
\textit{2T} & $72\%$ & 7.83 & 7.83 & 7.83 \cite{kraa+68tetrahedron} \\
\mr
\textit{3T} & $71\%$ & 7.32 & 7.33 & 7.38 \cite{filh+07ptrs}\\
\mr
\textit{F} & $79\%$ & 8.89 & 8.88 & 8.87 \cite{klap+90cjc}\\
\mr
\textit{2F} & $77\%$ & 7.65 & 7.66 & 7.90 \cite{nova+92jesrp} \\
\mr
\textit{3F} & $77\%$ & 7.16 & 7.16 & - \\
\mr
\textit{TF$_{ap}$} & $75\%$ & 7.68 & 7.72 & 7.85 \cite{nova+92jesrp} \\
\mr
\textit{TF$_{p}$} & $76\%$ & 7.69 & 7.69 & 7.85 \cite{nova+92jesrp} \\
\mr
\textit{TFT} & $73\%$ & 7.12 & 7.12 & - \\
\mr
\textit{FTF} & $74\%$ & 7.16 & 7.20 & - \\
\mr
\textit{TFTF} & $74\%$ & 6.89 & 6.89 & - \\
\mr
\textit{TFFT} & $73\%$ & 6.82 & 6.82 & - \\
\br
\end{tabular}
\end{indented}
\end{table}

We show in Table \ref{thio-furan_IP-alphaic}  the results adopting the PBE$\alpha_{ic}$ functional for a set of systems. The first column shows the Koopmans' IP values from PBE$\alpha_{ic}$, the second the results corrected by G0W0, and it can be seen that they are in good agreement with the experimental results included in the last column. Concerning the contribution of exact exchange in each case, as expected, from monomers to tetramers we also note that the internally-consistent $\alpha$ values are large for the monomers (shorter molecules), and go slightly smaller with increasing chain size.

\section{Discussion}

From our results above, we can see that for an isolated thio-furan oligomer the ordering of units is very relevant, that is, when we need planarity for each chain---and this can be relevant for the film structural stacking---nearest-neighboring of thiophene units should be avoided. As such, we suggest that the polymerization of chains following the prescription (..\textit{TFTF}..), that we name as oligo-thienylfuran OThFu from now on, may be very convenient. Indeed, a recent study\cite{yao+20mcf} showed that electrochemical polymerization of OThFu produces films with good flexibility and electrochemical activity, higher conductivity than polythiophene or polyfuran, and even good stability in air. As for the optical properties we see that the relative intensity of the first absorption, compared to the higher energy absorption peaks, increases with chain length from dimers to tetramers in a relevant trend. More important, regarding the solar radiance at the Earth surface, we find that the energy of this first transition for the OThFu chains reaches a convenient value already at tetramers. Finally, going on to the ionization potential of the isolated oligomers, our results for the tetramers are at IP$\sim 6.9$eV.

A very important property for the design of organic/inorganic solar cells is the IP of the organic moiety, since usually the molecule is deposited on a large-gap semiconductor surface. Then, in order to have photoexcited electron-transfer from the molecular system to the semiconductor, we need a good alignment of the IP and the acceptor band structure, in particular referring to the conduction band minimum CBM. The energy difference between the CBM and the IP should be such that, once the optical absorption by the organic moiety creates the electron-hole pair in the organic region, the electron can decay to the semiconductor leaving the hole in the original molecular system. Comparing the CBM energies of the {TiO$_2$} and {ZnO} semiconducting oxides, $E_{CBM}\sim 4.5 - 4.9$eV, with the tetramers first exciton energy at $E\sim 3$eV, we can conclude that the charge transfer organic-inorganic is energetically allowed. Longer chains might not be as convenient, since the absorption energy will probably decay so as not to allow efficient donor-acceptor transfer. In any case, the oxide specific surface termination, or the thickness of the oxide slab must be considered in order to guarantee a convenient alignment.  

\section{Conclusion}

In this work we present a theoretical study of structural, electronic and optical properties of thiophene and furan oligomers and co-oligomers relevant for photovoltaic devices.
We used ab-initio procedures for all properties, from DFT to Many-Body GW and Bethe-Salpeter equation, and studied different combinations and compositions of unit-neighbouring for thiophene \textit{T} and furan \textit{F} units. 

We find that segments defined by \textit{T-F} sequences (OThFu, explicity no \textit{T-T} neighborhood) show planar conformation, which can be useful for film deposition.  It is important to highlight that for tetramers the first absorption presents high intensity, and energy at the visible limit of solar radiance at the earth surface. Moreover, our results for OThFu tetramers show a convenient placement of the ionization potential relative to the conduction band of well known semiconducting oxides, which can lead to an efficient charge transfer from molecule to oxide. 

These combined properties point to a very interesting behaviour for OThFu, that can certainly be explored in the design of new systems for optoelectronic applications.\\

\ack

The authors thank the Brazilian agencies: CNPq (Scholarship process 131876/2020-1), and the National Institute for Science and Technology on Organic Electronics (CNPq 573762/2008-2 and FAPESP 2008/57706-4); and the University of São Paulo for access to the High Performance Computing Center (HPC-USP).\\

\bibliographystyle{iopart-num}
\bibliography{bastosetalBIBLIO-NAMES,bastosetalBIBLIO}

\providecommand{\newblock}{}
\begin{thebibliography}{10}
\expandafter\ifx\csname url\endcsname\relax
  \def\url#1{{\tt #1}}\fi
\expandafter\ifx\csname urlprefix\endcsname\relax\def\urlprefix{URL }\fi
\providecommand{\eprint}[2][]{\url{#2}}

\bibitem{kim+03nl}
Kim Y~G, Walker J, Samuelson L~A and Kumar J 2003 {\em Nano~Lett.~\/} {\bf 3}
  523--525

\bibitem{blum+08apl}
Blumstengel S, Koch N, Sadofev S, Sch{\"a}fer P, Glowatzki H, Johnson R, Rabe J
  and Henneberger F 2008 {\em Appl.~Phys.~Lett.~\/} {\bf 92} 169

\bibitem{sant+14jpcc}
Alves-Santos M, Jorge L~M~M, Caldas M~J and Varsano D 2014 {\em
  J.~Phys.~Chem.~C\/} {\bf 118} 13539--13544

\bibitem{bell+94jpc}
Belletete M, Leclerc M and Durocher G 1994 {\em J.~Phys.~C\/} {\bf 98}
  9450--9456

\bibitem{filh+07ptrs}
da~Silva~Filho D~A, Coropceanu V, Fichou D, Gruhn N~E, Bill T~G, Gierschner J,
  Cornil J and Bredas J~L 2007 {\em Phil.~Trans.~Royal~Soc.\/} {\bf 365}
  1435--1452

\bibitem{holl+14pccp}
Holland D, Trofimov A, Seddon E, Gromov E, Korona T, De~Oliveira N, Archer L,
  Joyeux D and Nahon L 2014 {\em Phys.~Chem.~Chem.~Phys.~\/} {\bf 16}
  21629--21644

\bibitem{bhag+21es}
Bhagat S, Leal W~D, Majewski M~B, Simbrunner J, Hofer S, Resel R and Salzmann I
  2021 {\em Electron.~Struct.~\/} {\bf 3} 034004

\bibitem{ramo-cald18arxiv}
Ramos R and Caldas M~J 2018 {\em arXiv preprint arXiv:1805.10335\/}

\bibitem{sney+21sa}
Sneyd A~J, Fukui T, Pale{\v{c}}ek D, Prodhan S, Wagner I, Zhang Y, Sung J,
  Collins S~M, Slater T~J, Andaji-Garmaroudi Z {\em et~al.\/} 2021 {\em
  Sci.~Adv.~\/} {\bf 7} eabh4232

\bibitem{gand11gc}
Gandini A 2011 {\em Green~Chem.\/} {\bf 13} 1061--1083

\bibitem{gidr+10jacs}
Gidron O, Diskin-Posner Y and Bendikov M 2010 {\em J.~Am.~Chem.~Soc.~\/} {\bf
  132} 2148--2150

\bibitem{gidr+11cc}
Gidron O, Dadvand A, Sheynin Y, Bendikov M and Perepichka D~F 2011 {\em
  Chem.~Commun.~\/} {\bf 47} 1976--1978

\bibitem{gidr-bend14ac}
Gidron O and Bendikov M 2014 {\em Angew.~Chem.~Int.~Ed.~\/} {\bf 53} 2546--2555

\bibitem{cao-rupa17cej}
Cao H and Rupar P~A 2017 {\em Chem.~Eur.~J.~\/} {\bf 23} 14670--14675

\bibitem{gidr+13cc}
Gidron O, Varsano N, Shimon L~J, Leitus G and Bendikov M 2013 {\em
  Chem.~Commun.~\/} {\bf 49} 6256--6258

\bibitem{xion+16am}
Xiong Y, Tao J, Wang R, Qiao X, Yang X, Wang D, Wu H and Li H 2016 {\em
  Adv.~Mater.~\/} {\bf 28} 5949--5953

\bibitem{yuny+16acsml}
Qiu Y, Fortney A, Tsai C~H, Baker M~A, Gil R~R, Kowalewski T and Noonan K~J
  2016 {\em ACS~Macro~Lett.~\/} {\bf 5} 332--336

\bibitem{stee+19jpcc}
Steen A~E, Ellington T~L, Nguyen S~T, Balasubramaniam S, Chandrasiri I, Delcamp
  J~H, Tschumper G~S, Hammer N~I and Watkins D~L 2019 {\em J.~Phys.~Chem.~C\/}
  {\bf 123} 15176--15185

\bibitem{yao+20mcf}
Yao W, Shen L, Liu P, Liu C, Xu J, Jiang Q, Liu G, Nie G and Jiang F 2020 {\em
  Mater.~Chem.~Front.~\/} {\bf 4}(2) 597--604

\bibitem{yao+21CJPS}
Yao W~Q, Liu P~P, Zhou W~Q, Duan X~M, Xu J~K, Yang J~J, Ming S~L, Li M and
  Jiang F~X 2021 {\em Chinese J. Polym. Sci.~\/} {\bf 39} 344--354

\bibitem{vali+10cpc}
Valiev M, Bylaska E~J, Govind N, Kowalski K, Straatsma T~P, Van~Dam H~J, Wang
  D, Nieplocha J, Apra E, Windus T~L {\em et~al.\/} 2010 {\em
  Comput.~Phys.~Commun.~\/} {\bf 181} 1477--1489

\bibitem{hohe+64pr}
Hohenberg P and Kohn W 1964 {\em Phys.~Rev.~\/} {\bf 136} B864

\bibitem{kohn+65pr}
Kohn W and Sham L~J 1965 {\em Phys.~Rev.~\/} {\bf 140} A1133

\bibitem{perd+96prl}
Perdew J~P, Burke K and Ernzerhof M 1996 {\em Phys.~Rev.~Lett.~\/} {\bf 77}(18)
  3865--3868

\bibitem{perd+96jcp}
Perdew J~P, Ernzerhof M and Burke K 1996 {\em J.~Chem.~Phys.~\/} {\bf 105}
  9982--9985

\bibitem{adam-baro99jcp}
Adamo C and Barone V 1999 {\em J.~Chem.~Phys.~\/} {\bf 110} 6158--6170

\bibitem{dunn89jcp}
Dunning~Jr T~H 1989 {\em J.~Chem.~Phys.~\/} {\bf 90} 1007--1023

\bibitem{raos+03cpl}
Raos G, Famulari A and Marcon V 2003 {\em Chem.~Phys.~Lett.~\/} {\bf 379}
  364--372

\bibitem{bloo+14jctc}
Bloom J~W and Wheeler S~E 2014 {\em J.~Chem.~Theory ~Comput.~\/} {\bf 10}
  3647--3655

\bibitem{blum+09cpc}
Blum V, Gehrke R, Hanke F, Havu P, Havu V, Ren X, Reuter K and Scheffler M 2009
  {\em Comput.~Phys.~Commun.~\/} {\bf 180} 2175--2196

\bibitem{liu+20jcp}
Liu C, Kloppenburg J, Yao Y, Ren X, Appel H, Kanai Y and Blum V 2020 {\em
  J.~Chem.~Phys.~\/} {\bf 152} 044105

\bibitem{zhan+13njp}
Zhang I~Y, Ren X, Rinke P, Blum V and Scheffler M 2013 {\em New J.~Phys.~\/}
  {\bf 15} 123033

\bibitem{hirs77tca}
Hirshfeld F~L 1977 {\em Theor.~Chem.~Acta\/} {\bf 44} 129--138

\bibitem{jane-pete11cms}
Murray J~S and Politzer P 2011 {\em Comp.~Mater.~Sci.~\/} {\bf 1} 153--163

\bibitem{bond64jpc}
Bondi A 1964 {\em J.~Phys.~C\/} {\bf 68} 441--451

\bibitem{jmol-jmol13}
Jmol 2013 {\em Jmol web page: http://www. jmol. org/\/}

\bibitem{hedi65pr}
Hedin L 1965 {\em Phys.~Rev.~\/} {\bf 139} A796

\bibitem{hybe-loui85prl}
Hybertsen M~S and Louie S~G 1985 {\em Phys.~Rev.~Lett.~\/} {\bf 55} 1418

\bibitem{godb+86prl}
Godby R, Schl{\"u}ter M and Sham L 1986 {\em Phys.~Rev.~Lett.~\/} {\bf 56} 2415

\bibitem{rohl-loui00prb}
Rohlfing M and Louie S 2000 {\em Phys.~Rev.~B\/} {\bf 62} 4927--4944

\bibitem{pinh+15prb}
Pinheiro~Jr M, Caldas M~J, Rinke P, Blum V and Scheffler M 2015 {\em
  Phys.~Rev.~B\/} {\bf 92} 195134

\bibitem{perk+21JPCA}
Perkins M~A, Cline L~M and Tschumper G~S 2021 {\em J.~Phys.~Chem.~A\/} {\bf
  125} 6228--6237

\bibitem{garc-karp09cpl}
Sancho-Garc{\'\i}a J~C and Karpfen A 2009 {\em Chem.~Phys.~Lett.~\/} {\bf 473}
  49--56

\bibitem{samd+93sm}
Samdal S, Samuelsen E~J and Volden H~V 1993 {\em Synth.~Met.~\/} {\bf 59}
  259--265

\bibitem{vera+74jacs}
Bucci P, Longeri M, Veracini C~A and Lunazzi L 1974 {\em J.~Am.~Chem.~Soc.~\/}
  {\bf 96} 1305--1309

\bibitem{pach+20jmcc}
Pachariyangkun A, Suda M, Hadsadee S, Jungsuttiwong S, Nalaoh P,
  Pattanasattayavong P, Sudyoadsuk T, Yamamoto H~M and Promarak V 2020 {\em
  J.~Mater.~Chem.~C\/} {\bf 8}(48) 17297--17306

\bibitem{palm+95cp}
Palmer M~H, Walker I~C, Ballard C~C and Guest M~F 1995 {\em Chem.~Phys.~\/}
  {\bf 192} 111--125

\bibitem{hung+16prb}
Hung L, Felipe H, Souto-Casares J, Chelikowsky J~R, Louie S~G and
  {\"O}{\u{g}}{\"u}t S 2016 {\em Phys.~Rev.~B\/} {\bf 94} 085125

\bibitem{prlj+15jpcl}
Prlj A, Curchod B~F, Fabrizio A, Floryan L and Corminboeuf C 2015 {\em
  J.~Phys.~Chem.~Lett.\/} {\bf 6} 13--21

\bibitem{schr+08jcp}
Schreiber M, Silva-Junior M~R, Sauer S~P~A and Thiel W 2008 {\em
  J.~Chem.~Phys.~\/} {\bf 128} 134110

\bibitem{gomb+04jpsj}
Gombojav B, Yoshinari T, Itoh H, Nagasaka S~i, Kuriyama Y and Koyama K 2004
  {\em J.~Phys.~Soc.~Jpn.~\/} {\bf 73} 3166--3170

\bibitem{diaz+81jecie}
Diaz A, Crowley J, Bargon J, Gardini G and Torrance J 1981 {\em
  J.~Electroanal.~Chem.~Interfacial~Electrochem.\/} {\bf 121} 355--361

\bibitem{wynb+71joc}
Wynberg H, Sinnige H and Creemers H 1971 {\em J.~Org.~Chem.~\/} {\bf 36}
  1011--1013

\bibitem{grig+66jcsc}
Grigg R, Knight J and Sargent M 1966 {\em J.~Chem.~Soc.~C\/}  976--981

\bibitem{klas+82jcs}
Klasinc L, Sablji{\'c} A, Kluge G, Rieger J and Scholz M 1982 {\em
  J.~Chem.~Soc.{,}~Perkin~Trans.~2\/} (5) 539--543

\bibitem{kraa+68tetrahedron}
Kraak A and Wynberg H 1968 {\em Tetrahedron~Lett.~\/} {\bf 24} 3881--3885

\bibitem{klap+90cjc}
Klapstein D, MacPherson C~D and O'Brien R~T 1990 {\em Can.~J.~Chem.\/} {\bf 68}
  747--754

\bibitem{nova+92jesrp}
Novak I, Ng S, Chua Y, Mok C and Huang H 1992 {\em
  J.~Electron~Spectrosc.~Relat.~Phenom.~\/} {\bf 61} 143 -- 147 ISSN 0368-2048

\end{thebibliography}
\pagebreak

\end{document}